\documentclass[doublecolumn]{IEEEtran}

\usepackage{amsmath, amsthm, amsfonts, amssymb, amsbsy,nccmath}
\usepackage{mathtools}
\usepackage{bbm}
\usepackage{acronym}  
\usepackage[dvips]{color}
\usepackage{epsf}
\usepackage{times}
\usepackage{epsfig}
\usepackage{notoccite}
\usepackage{graphicx}
\usepackage{epstopdf}
\usepackage{pstricks}
\usepackage{amssymb}
\usepackage{amsxtra}
\usepackage{here}
\usepackage{rawfonts}
\usepackage{float}
\usepackage{times}
\usepackage{url}
\usepackage{cite}
\usepackage{array, makecell} %

\usepackage{caption}
\usepackage{subcaption}
\usepackage{algorithm}
\usepackage{algpseudocode}
\usepackage{blindtext}
\usepackage{enumitem}
\usepackage{xcolor,cite,etoolbox}
\usepackage{relsize}
\usepackage{lipsum}
\usepackage{graphicx}
\usepackage{tabularx}
\usepackage{xparse}
\usepackage{array}
\newcolumntype{P}[1]{>{\centering\arraybackslash}p{#1}}
\usepackage{mleftright}

\usepackage{mathtools}

\usepackage{graphics}
\usepackage{physics}
\usepackage{amssymb}
\usepackage{siunitx}
\include{newcommands}
\usepackage{multicol}

\usepackage[nomain,acronym,shortcuts]{glossaries}
\makeglossaries
\newcommand*{\acro}[3][]{\newacronym[#1]{#2}{#2}{#3}}

\acro{D2D}{device-to-device}
\acro{SC}{semantic communication}
\acro{SIR}{signal-to-interference-ratio}
\acro{SINR}{signal-to-interference-plus-noise-ratio}
\acro{PCP}{Poisson cluster process}
\acro{AD}{algorithm distillation}
\acro{NN}{neural network}
\acro{PRB}{physical resource block}
\acro{CoMP}{coordinated multi-point}
\acro{BS}{base station} 
\acro{MD-CoMP}{macrodiversity CoMP transmission}
\acro{MAC}{medium-access-control}
\acro{RAN}{radio access network}
\acro{JT-CoMP}{joint transmission CoMP}
\acro{CoMP-JT}{coordinated multipoint joint transmission}
\acro{SBS}{small base station}
\acro{MDSD}{multiple devices to a single device}
\acro{MDS}{maximum distance separable}
\acro{SCN}{small cell network}
\acro{PPP}{Poisson point process}
\acro{TCP}{Thomas cluster process}
\acro{CSI}{channel state information}
\acro{PDF}{probability distribution function}
\acro{PMF}{probability mass function}
\acro{RV}{random variable}
\acro{i.i.d.}{independently and identically distributed}
\acro{MBMS}{multimedia broadcasting multicasting service}
\acro{EE}{energy efficiency}
\acro{HCP}{hard-core placement}
\acro{CCDF}{complementary cumulative distribution function}
\acro{CDF}{cumulative distribution function}
\acro{PC}{probabilistic caching}
\acro{RC}{random caching}
\acro{CPF}{caching popular files} 
\acro{PGFL}{probability generating functional}
\acro{KKT}{Karush-Kuhn-Tucker}
\acro{PGF}{point generating function}
\acro{SCA}{successive convex approximation}
\acro{HD}{high-definition}
\acro{OSI}{open systems interconnection}
\acro{FHD}{full-high-definition}
\acro{UHD}{ultra-high-definition}
\acro{VR}{virtual reality}
\acro{AR}{augmented reality}
\acro{5G}{fifth-generation}
\acro{QoS}{quality-of-service}
\acro{QoE}{quality-of-experience}
\acro{IoT}{internet of things}
\acro{MHCPP}{Matern hardcore point process}
\acro{LoS}{line-of-sight}
\acro{NLoS}{non-line-of-sight}
\acro{PSD}{power spectral density}
\acro{MEC}{mobile edge computing}
\acro{E2C}{edge-to-cloud}
\acro{E2E}{end-to-end}
\acro{THz}{terahertz}
\acro{CLT}{central limit theorem}
\acro{HQ}{High Quality}
\acro{eMBB}{enhanced mobile broadband}
\acro{URLLC}{ultra reliable low latency communications}
\acro{mmWave}{millimeter wave}
\acro{EVT}{extreme value theory}
\acro{GEV}{generalized extreme value}
\acro{LIS}{large intelligent surface}
\acro{RIS}{reconfigurable intelligent surface}
\acro{RF}{radio frequency}
\acro{UE}{user equipment}
\acro{MIMO}{multiple-input multiple-output}
\acro{EVaR}{entropic value-at-risk}
\acro{DNN}{deep neural network}
\acro{MDP}{Markov decision process}
\acro{RL}{reinforcement learning}
\acro{RNN}{recurrent neural network}
\acro{GNN}{graph neural network}
\acro{ANN}{artificial NN}
\acro{LSTM}{long short-term memory}
\acro{ReLu}{rectified linear unit}
\acro{VaR}{value-at-risk}
\acro{SNR}{signal-to-noise ratio}
\acro{AoSA}{array of subarray}
\acro{XR}{extended reality}
\acro{AoA}{angle of arrival}
\acro{ULA}{uniform linear array}
\acro{AoD}{angle of departure}
\acro{EM}{electromagnetic}
\acro{HRLLC}{s high-rate and high-reliability low latency communications}
\acro{6DoF}{six degrees of freedom}
\acro{MR}{mixed reality}
\acro{PAPR}{peak to average power ratio}
\acro{OFDM}{orthogonal frequency-division multiplexing}
\acro{OFDMA}{orthogonal frequency-division multiple access}
\acro{SC-FDM}{single carrier frequency-division multiplexing}
\acro{ToA}{time of arrival}
\acro{MUSIC}{multiple signal classification}
\acro{IoE}{Internet of Everything}
\acro{DT}{digital twin}
\acro{PT}{physical twin}
\acro{CT}{cyber twin}
\acro{DRL}{deep reinforcement learning}
\acro{FL}{federated learning}
\acro{DL}{deep learning}
\acro{CRAS}{connected robotics and autonomous system}
\acro{CL}{continual learning}
\acro{BF}{beamforming}
\acro{MSE}{mean squared error}
\acro{EWC}{elastic weight consolidation}
\acro{ML}{machine learning}
\acro{GD}{gradient descent}
\acro{MLP}{multi layer perceptron}
\acro{TL}{transfer learning}
\acro{AI}{artificial intelligence}
\acro{NFT}{non fungible token}
\acro{H2A}{human-to-avatar}
\acro{A2A}{avatar-to-avatar}
\acro{UAV}{unmanned aerial vehicle}
\acro{NTN}{non-terrestrial network}
\acro{ISAC}{integrated sensing and communication }
\acro{CIS}{connected intelligence system}
\acro{QoVE}{quality of virtual experience}
\acro{OOD}{out-of-distribution}
\acro{XAI}{explainable AI}
\acro{LLM}{large language model}
\acro{CNN}{convolutional neural network}
\acro{KPI}{key performance indicator}
\acro{SCM}{structural causal model}
\acro{CGM}{causal graphical model}
\acro{DAG}{directed acyclic graph}
\acro{IIT}{integrated information theory}
\acro{CI}{connected intelligence}
\acro{UL}{uplink}
\acro{KF}{Kalman filter}
\acro{GAN}{generative adversarial network}
\acro{6G}{sixth generation}
\acro{MCMC}{Markov chain Monte-Carlo}
\acro{MAB}{multi-armed bandit}

\usepackage{datetime}
\usepackage{amssymb}
\usepackage{subcaption}
\usepackage{verbatim}

\theoremstyle{definition}
\newtheorem{definition}{Definition}

\newcommand{\beq}{\begin{equation}}
\newcommand{\eeq}{\end{equation}}

\newcommand{\mD}{\mbox{$\mathcal D$}}
\newcommand{\mP}{\mbox{$\mathcal P$}}

\newcommand{\bmX}{\boldsymbol{X}}

\DeclareMathOperator*{\argmax}{arg\,max}

\newcommand{\mI}{\mathcal{I}}

\newcommand{\mF}{\mathcal{F}}

\newcommand{\mA}{\mathcal{A}}
\newcommand{\mE}{\mathcal{E}}

\newcommand{\mV}{\mathcal{V}}

\newcommand{\mM}{\mathcal{M}}

\newcommand{\mU}{{\mathcal U}}

\def\adots{\mathinner{\mskip0mu\raise0pt\vbox{\kern7pt\hbox{.}}\mskip3mu
          \raise4pt\hbox{.}\mskip3mu\raise8pt\hbox{.}\mskip0mu}}

\newcommand{\bmc}{{\boldsymbol c}}

\newcommand{\bmh}{\bfh}

\newcommand{\mL}{\mathcal{L}}
\newcommand{\mC}{\mathcal{C}}

\newcommand{\mG}{\mathcal{G}}

\usepackage{bm}

\newcommand{\bmx}{{\bm x}}
\newcommand{\bmy}{{\bm y}}
\newcommand{\bmv}{{\bm v}}
\newcommand{\bmo}{{\bm o}}

\newcommand{\bmu}{{\bm u}}

\newcommand{\mbE}{{\mathbb E}}

\renewcommand{\bmh}{{\bm h}}

\newcommand{\bmY}{{\bm Y}}
\newcommand{\bms}{{\bm s}}

\newcommand{\bmA}{{\bm A}}
\newcommand{\bmr}{{\bm r}}

\newcommand{\bmD}{{\boldsymbol {D}}}

\newcommand{\bma}{{\bm a}}

\newcommand{\bmU}{\bm U}

\newcommand{\btheta}{\boldsymbol{\theta}}

\newcommand{\bphi}{\boldsymbol{\phi}}

\begin{document}
\title{Causal Reasoning: Charting a Revolutionary Course for Next-Generation AI-Native Wireless Networks\vspace{-6mm}
	\author{Christo Kurisummoottil Thomas, \emph{Member, IEEE,} Christina Chaccour, \emph{Member, IEEE,}\\
		Walid Saad,  \emph{Fellow, IEEE,} 
 Mérouane Debbah, \emph{Fellow, IEEE}, and Choong Seon Hong, \emph{Senior Member, IEEE}}
	\thanks{C. Thomas, and W. Saad are with the Wireless@VT, Bradley Department of Electrical and Computer Engineering, Virginia Tech, Arlington, VA, USA. 
(Emails: \protect{christokt@vt.edu}, \protect{walids@vt.edu}).}
    \thanks{C. Chaccour is with Ericsson, Inc., Plano, Texas, USA, Email: \protect{christina.chaccour@ericsson.com}.}
	\thanks{ M. Debbah is with Khalifa University of Science and Technology in Abu Dhabi and also with CentraleSupelec, University Paris-Saclay, 91192 Gif-sur-Yvette, France, (Email: \protect{merouane.debbah@ku.ac.ae}).} 
 \thanks{C. S. Hong is with the Department of Computer Science and Engineering, School
of Computing, Kyung Hee University, Yongin-si
17104, Republic of Korea, (Email: \protect{cshong@khu.ac.kr})}
}

\maketitle
\vspace{-20mm}
\begin{abstract}
Despite the basic premise that next-generation wireless networks (e.g., 6G) will be artificial intelligence (AI)-native, to date, most existing efforts remain either qualitative or incremental extensions to existing ``AI for wireless'' paradigms. Indeed, creating AI-native wireless networks faces significant technical challenges due to the limitations of data-driven, training-intensive AI. These limitations include the black-box nature of the AI models, their curve-fitting nature, which can limit their ability to reason and adapt, their reliance on large amounts of training data, and the energy inefficiency of large neural networks. In response to these limitations, this article presents a comprehensive, forward-looking vision that addresses these shortcomings by introducing a novel framework for building AI-native wireless networks; grounded in the emerging field of causal reasoning. Causal reasoning, founded on causal discovery, causal representation learning, and causal inference, can help build explainable, reasoning-aware, and sustainable wireless networks. Embracing such a human-like AI foundation can revolutionize the design of AI-native wireless networks, laying the foundations for creating self-sustaining networks that ensure uninterrupted connectivity. Towards fulfilling this vision, we first highlight several wireless networking challenges that can be addressed by causal discovery and representation, including ultra-reliable beamforming for terahertz (THz) systems, near-accurate physical twin modeling for digital twins, training data augmentation, and semantic communication. We showcase how incorporating causal discovery can assist in achieving dynamic adaptability, resilience, and cognition in addressing these challenges. Furthermore, we outline potential frameworks that leverage causal inference to achieve the overarching objectives of future-generation networks, including intent management, dynamic adaptability, human-level cognition, reasoning, and the critical element of time sensitivity. We conclude by offering recommendations shaping the roadmap toward causality-driven AI-native next-generation wireless networking. 
\end{abstract}\vspace{-1mm}

\section{Introduction}
\label{Introduction}
\begin{figure*}[t]
\centerline{\includegraphics[width=\textwidth]{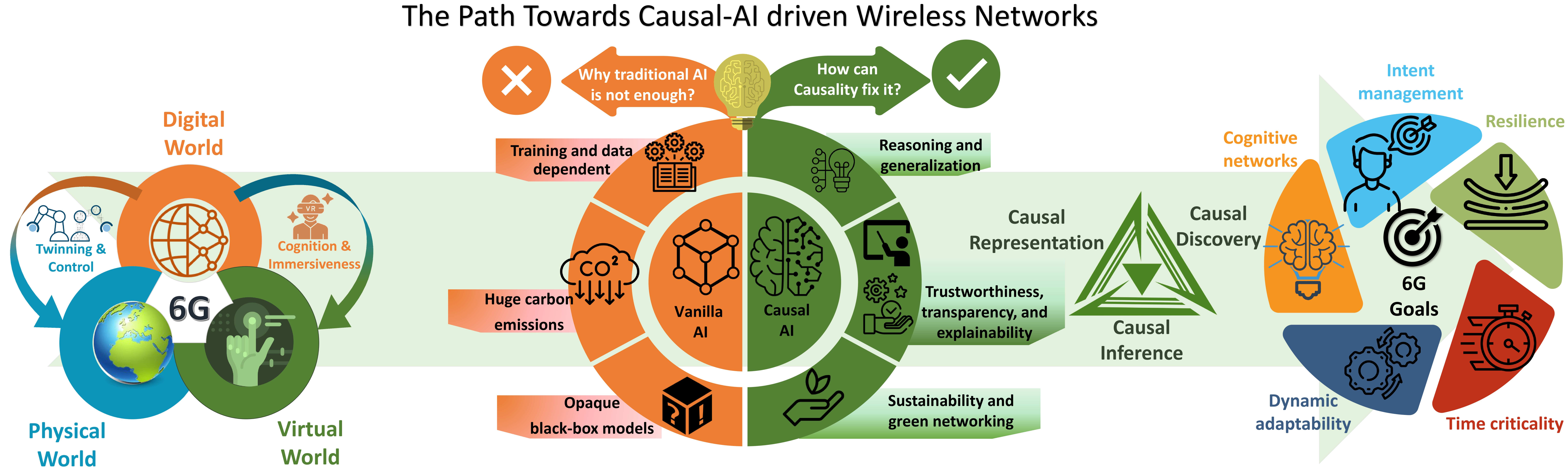}}\vspace{-1mm}
\caption{\small Causal AI vision for next-generation wireless networks. }
\label{CausalAI_Vision}\vspace{-0mm}
\vspace{-3.5mm}
\end{figure*}\vspace{-2mm}

As we peer into the future of next-generation wireless networks (e.g., 6G and beyond), the incorporation of \ac{AI} across the protocol stack of wireless systems is no longer just a theoretical concept—it is becoming a tangible reality \cite{XLinArxiv2023}. \emph{\ac{AI}-native} wireless systems must leverage \ac{ML} and \ac{AI} algorithms to design, optimize, and operate various aspects of the wireless system, including transceiver design, resource allocation, interference management, and many more. However, developing \ac{AI} native 6G wireless systems necessitates the design of novel AI frameworks that are tailored to several unique challenges of wireless systems: 1) \emph{dynamic adaptability} to ensure rapid adjustments to changing network conditions, user demands, and other environmental factors; 2) \emph{time criticality}, whereby 6G systems must deliver ultra-low latency and unwavering reliability, particularly for applications demanding split-second responsiveness; 3) \emph{intent management}, enabling networks to autonomously translate high-level business intents into network configurations in a closed-loop fashion, ensuring intent assurance across the network while maintaining overall network reliability;
4) \emph{resilience}, enabling 6G networks to withstand disruptions and maintain connectivity even in challenging scenarios; 5) \emph{non-linear signal dynamics}, that must be properly modeled to accurately capture the time-varying nature of multi-modal wireless signals, e.g., audio, video, haptics and olfactory signals; and 6) \emph{human-level cognition and reasoning} that must be integrated into the wireless system for intelligent decision-making, self-optimization, and efficient resource management, ultimately enabling both sample efficiency and communication efficiency, thereby achieving more with less data and resources. 

\indent Despite significant progress in AI-native wireless networks, existing solutions (e.g., \cite{ZhangCST2019}) remain predominantly data-driven and based on statistical learning. 
These traits pose certain challenges in overcoming the aforementioned challenges. 
Firstly, statistical AI models require \emph{extensive data volumes}, to ensure reliable performance. However, obtaining extensive datasets is not always feasible in wireless networks. Secondly, current AI model parameters are closely linked to their training data, requiring \emph{frequent and recurrent retraining} to maintain effectiveness as the wireless and \ac{RAN} environments undergo rapid changes. 
This consumes substantial communication and computational resources. The resulting training overhead could disrupt the communication phase, consequently impacting the stringent requirements of emerging wireless services, in terms of uninterrupted availability and near-zero end-to-end latency. Furthermore, existing AI tools frequently encounter challenges when it comes to reasoning, particularly when tasked with producing novel insights or combining already-seen experiences to generate fresh combinations or make logical deductions. 
Finally, the reliance of statistical AI on complex \acp{NN} and often opaque architectures makes it challenging to identify the root causes of any performance deviations or network operation issues at the granular level across the network stack, interfaces, and the profusion of devices and sensors. 
\vspace{-1mm}

To overcome these limitations, as envisioned in Fig.~\ref{CausalAI_Vision}, it is imperative to design new AI frameworks that: 1) enable \emph{reasoning and generalization} by exhibiting consistent performance across diverse wireless data distributions (inductive reasoning), as well as an ability to combine already seen experiences to generate novel logical conclusions (deductive reasoning), 2) possess the properties of \emph{explainability and transparency} which means they can operate reliably while providing insights into how and why they make specific decisions, thereby developing a deeper understanding of their actions across the wireless network protocol stack, and 3) promote the advancement of \emph{sustainable and green networks} by improving sample efficiency and creating more lightweight \ac{ML} models. Several methodologies can facilitate the development of generalizable AI, such as transfer learning, meta-learning, domain adaptation, and continual learning \cite{ChenLiuMorgan2018}. However, these approaches still require extensive and diverse datasets for efficacy and typically rely on \acp{NN} that lack transparency and interpretability. An alternative involves gathering extensive wireless data, similar to \acp{LLM} like ChatGPT. This data-driven networking approach, using \emph{foundation models} \cite{LeACM2022}, enhances AI models' adaptability in wireless contexts. However, generative AI models like ChatGPT, while useful as writing assistants, may introduce spurious data points, impacting wireless network resilience and causing faulty predictions. Recent studies, such as~\cite{AiyappaArxiv2023}, reveal increased errors in ChatGPT due to challenges like model collapse and catastrophic forgetting. Deploying ChatGPT-like AI models in mission-critical and time-critical settings, such as autonomous vehicles or digital twins of industrial plants, may pose risks. Additionally, large \ac{NN} architectures demand significant processing, leading to energy-inefficient networking, contradicting the pursuit of \emph{energy-conscious sustainability} objectives. 
\vspace{-0.3mm}

Essentially, if we are to efficiently design AI-native wireless networks, we must create AI framework that excel in performance and in emulating \emph{human-like reasoning capabilities}. ``Human-like reasoning capabilities" refer to the ability of \ac{AI} systems to think, make decisions, and understand information analogously  to how humans do, encompassing common sense, reasoning, contextual understanding, adaptability, as well as ethical and explanatory capabilities. 
Such models could alleviate the challenges associated with conventional \ac{ML} approaches by providing a deeper understanding of their decision-making processes and achieving superior generalization across an array of tasks and scenarios as required by next-generation AI-native wireless networks. Here, a promising avenue lies in the realm of \emph{causal reasoning} \cite{PearlBasic2018} and \cite{ScholkopfPIEEE2021}.

Causal reasoning involves recognizing cause-and-effect relations within the wireless data, helping the AI models understand how changes in one component of the system can impact or predict outcomes in other components, and, facilitating informed decision-making and effective problem-solving. 
Causality is realized through the processes of \emph{causal discovery, representation, and inference}. 
Causal discovery enables the creation of \emph{explainable AI-native wireless models}, allowing us to describe the network decisions that result in specific outcomes unlike the opacity of transfer or meta or continual learning schemes. Meanwhile, causal inference and representation facilitate human-like reasoning by leveraging concepts like \emph{interventions and counterfactuals}, which can be performed either online or offline on the learned causal model. In causal \ac{ML}, ``interventions" involve actively manipulating one or more network variables to observe the effects on other variables. ``Counterfactuals" refer to hypothetical scenarios where we assess what would have happened if certain interventions or changes had or had not occurred, providing insights into the causal impact of those interventions. 
Causal inference empowers wireless systems to extend the applicability of their predictions and decision-making capabilities across various data distributions (\emph{generalizability}), assuming a consistent data generation process (\emph{causal understanding}). This results in more resilient and efficient knowledge transfer across domains compared to transfer learning, enhanced sample efficiency (leading to \emph{sustainability}) compared to meta-learning, and dynamic adaptability to varying task or data distributions without the risk of catastrophic forgetting associated with continual learning. 
As shown in Fig.~\ref{CausalAI_Vision}, advances in designing causal reasoning frameworks will allow the development of AI-native wireless networks characterized by the previously discussed attributes that include reasoning awareness, intent management, resilience, dynamic adaptability, and real-time synchronization which align with the objectives of next-generation wireless networks, namely 6G and beyond.

\vspace{-2mm}\subsection{Contributions}

The main contribution of this article is a novel and holistic vision that articulates fundamental principles necessary
to build next-generation causality-driven, AI-native wireless networks. Our key contributions include:

\begin{itemize}
\item We first present the fundamental building blocks of causal reasoning, and motivate the need for incorporating them in future wireless networks through a simple example.
\item We identify the inherent challenges in current AI-native wireless systems, like lack of explainability, limited reasoning capability, and energy inefficiency. We then discuss the potential of causal reasoning for addressing these challenges.
\item We demonstrate that causal discovery and representation accelerate the development of interpretable and cognition-enhanced wireless systems. This is achieved by decoding causal relationships within wireless data as a graph and analyzing cause-and-effect dynamics using \acp{CGM}. These insights support use cases like dynamic channel tracking in THz systems, precise \ac{DT} modeling, training data augmentation, the creation of resilient wireless networks, and minimalist representation for \ac{SC} systems. 
\item 
We introduce a robust analytical framework based on causal inference for addressing wireless control problems. This framework leverages the learned \acp{CGM} and incorporates causal \ac{RL}, causal \ac{MAB}, and causal multi-objective optimization. We further outline how causality can address challenges related to partial observability, scalability, and real-time communication protocol design in \ac{ISAC} systems. 
Leveraging interventional and counterfactual reasoning within causal \ac{ML}, these frameworks aim to achieve the wireless goals of dynamic adaptability, intent management, and time criticality.  
\end{itemize}

 
\section{Causality primer}
  \label{Primer}

Our first key step is to fundamentally explain why and how wireless models can benefit from causal structure learning. Towards this aim, this section provides a primer on causality, exploring its fundamental concepts and its relevance to wireless systems. 
\begin{figure*}[t]
\centerline{\includegraphics[width=5.8in,height=1.5in]{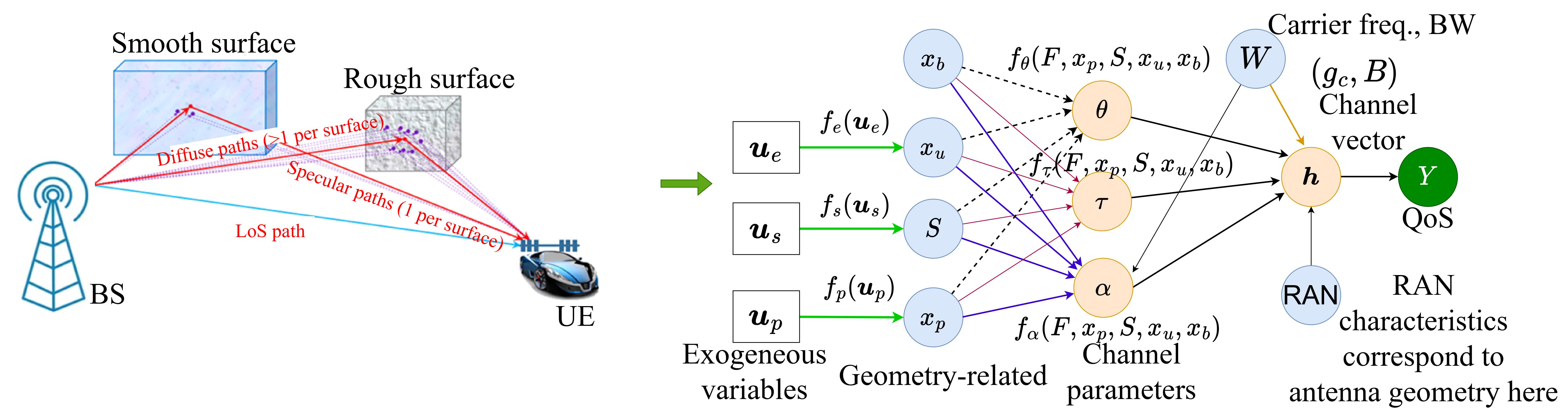}}\vspace{-0.5mm}
\caption{\small Mapping a propagation environment to a causal diagram. For simplicity, the diagram shows the causal dependencies for a single path, while the discussions in the paper is revolved around a more general case, that encompassed multiple scattering elements and multiple channel paths.}
\label{Causal_wireless_eg}\vspace{-0mm}
\vspace{-6mm}
\end{figure*}
\vspace{-4mm}\subsection{Why do Statistical ML Models Fall Short in Reasoning over Wireless Networks?}
\label{Primer_why}

Causality provides insights into the actual cause-and-effect dynamics within a system, akin to the ``physics" of the system, and hence goes beyond mere correlation or statistical dependency. Consider a simple wireless beam training example for a \ac{THz} system where pure statistical deep learning methods may fail. In 
this example, the system is provided with a dataset containing channel or network parameters $(X)$ and their corresponding labels $(Y)$ that represent a beam index. The model's objective is to learn a mapping function, $Y = f(X)$, that accurately predicts the labels based on the observed channel data. In this purely statistical approach, the model's learning process relies on identifying and capturing statistical dependencies (in short, the joint distribution $P(X,Y)$ between input $(X)$ and output $(Y)$ variables. To illustrate why a pure statistical approach is insufficient, we expand the simple beam tracking example as illustrated in Fig.~\ref{Causal_wireless_eg}. The physics behind the generation of the channel geometry encompassing location parameters and scatter parameters (reflection coefficient $S$) are represented by variables $\bmu_e,\bmu_s,$ and $\bmu_p$, that are exogenous to the channel model. The dependency of these exogenous variables on the propagation geometry parameters via $f_e, f_s,$ and $f_p$, cannot be accurately determined. As shown in Fig.~\ref{Causal_wireless_eg}, the angle of arrival (AoA) $\theta$, delay $\tau$, and path amplitudes $\alpha$, are generated using either deterministic or random functions \cite{TseCUP2005} $\theta = f_{\theta}(F,x_b,S,x_u,x_p)$ and $\tau = f_{\tau}(x_b,x_u,S,x_p)$. $x_p$ and $\alpha$ represent the position, and amplitude corresponding to distinct scatterers.  $x_b$ and $x_u$ are the locations of \ac{BS}, and \ac{UE}, respectively.
Conventional \ac{ML} solutions approximate the non-linear channel function, given the extracted channel parameters from the training data. The \ac{BS} employs received pilot observations and \ac{UE} feedback during the \ac{UL} as their ML training data. 
 In THz bands, increased penetration losses and reduced reflection coefficients heighten vulnerability to signal blockages, endangering transmission reliability \cite{ChaccourITJ2022}. A sudden obstruction, such as a moving object or a person, disrupts the \ac{LoS} path between the \ac{UE} and \ac{BS}. 
 The resulting channel exhibits substantial power fluctuations, far more pronounced than those encountered in mmWave or sub-6 GHz bands. Classical \ac{ML} models could handle smaller variations in non-linear channel models, however, they cannot capture the immense dynamic variations typical of THz frequencies.  This limitation arises from their inherent curve-fitting nature, where the \ac{NN} parameters are closely linked to the specific dataset used for their training. 
 Furthermore, such black-box models cannot explain the rationale behind the selection of a specific beam index for a given set of channel parameters. 
 Hence, a fundamental question arises: \emph{how can we transition to a AI model that can adapt its performance to various channel distributions, be resilient to unforeseen scenarios, and provide interpretable results, unlike black-box models?} One answer is through the use of causal reasoning \cite{PearlBasic2018}. In causal \ac{ML}, the system can obtain an understanding of the causal relationships that lead to the generation of the channel as well as the relations among network variables that determine the user \ac{QoE}. These causal relationships can be captured using directed edges in a graph (see Fig.~\ref{Causal_wireless_eg}). 
 Understanding these dependencies enables AI inference engines to manipulate them in real-time, thereby 1) enhancing the network's predictive capabilities; and 2) ensuring better network decisions aligned with long-term goals, promoting seamless network connectivity. Next, we look at how grasping these causal relationships confers advantages over traditional deep learning-based wireless systems.

\vspace{-3mm}
 \subsection{How Can We Embed Causality Awareness into AI-Native Wireless Models?}

In our example, the wireless environment is represented by $X = \{x_b,x_u,x_p, S\}$ that encompasses factors like the physical object layout, which influences channel conditions, and users' locations. Components of $X$ are represented by the possibly non-linear functions $f_e,f_s,$ and $f_p$. 
Here, to assess the influence of $X$ on $Y$, in addition to utilizing offline data such as \ac{UL} channel measurements, a causality-based framework integrates online data generated from various experiments using the causal structure. These experiments, also called interventions, can be defined as picking the functions $f_i, \forall i \in \{e,s,p,\theta,\tau,\alpha\}$ from a set $\mF_i$ of different environment as well as \ac{RAN} characteristics. Using interventions, system designers can understand how different factors within the wireless environment, represented by $X$, influence the \ac{QoE} experienced by end users, represented by $Y$. 
Contrary to this, the best Classical AI-native systems can do is generate extensive set of random data distributions to train the AI model, resulting in sample inefficiency. This example illustrates the potential implications of causality in wireless systems. 
\renewcommand{\arraystretch}{3.0}
\begin{table*}\fontsize{14pt}{14pt}\selectfont
\begin{center}
 \resizebox{\textwidth}{!}{%
 \begin{tabular}{ c|c|c|c|c }  
 \hline
\textbf{Layer
(Symbolic)}
 & \textbf{Typical
Activity} & \textbf{Typical
Question} & \textbf{Wireless Network Example} & \textbf{AI Model} \\   \hline
 \!\!\!\!\!\!$\mL_1:$ Associational $p(y\mid x)$ & \makecell{\emph{\textbf{Seeing}} \\ (e.g., \ac{UL} channel measurements, \\ multi-modal sensors for \ac{DT})} & \makecell{What is?
How would seeing
$X$ \\ change my belief
in $Y$?} & \makecell{What does channel quality \\ tells us about \ac{QoE} at the user?} &  \makecell{Supervised and
Unsupervised
Learning, \\ variational auto-encoders, \\ \ac{RNN},\\ transformers, large lanugage models.}\\ \hline
\hspace{6mm}\!\!\!\!\!\!$\mL_2:$ \hspace{1mm}Interventional $p(y\mid {do}(x))$& \makecell{\emph{\textbf{Doing}} \\ (e.g., dynamic network slicing decisions, \\ dynamic mobile-edge computing \\ offloading decisions)} & \makecell{What if? \\
What if I do $X$?}
  & \makecell{What if I increase  \\ the number of antennas, \\ would my BF quality be adjusted\\  for better user QoE?} & \makecell{Causal \ac{RL}, \\ causal multi-armed bandit.}  \\ \hline
\hspace{6mm} \!\!\!\!\!\!$\mL_3:$ \hspace{1mm}Counterfactual $p(y_x\mid x^{\prime},y^{\prime})$& \makecell{\emph{\textbf{Imagining}} \\ (e.g., altering network infrastructure, \\ transmission schemes for\\ recuperating from malicious firmware \\ attack that corrupt the AI algorithms)} & \makecell{Why?
What if I\\ had acted
differently?} & \makecell{Was it the frequency band, \\ rather than the number of antennas, \\ that had a detrimental effect on user QoE?} & \makecell{Theory of mind based reasoning \cite{YuanArxiv2021}, \\ counterfactual explanation policies \\ using RL\cite{MadumalAAAI2020}.}\\ 
 \hline
\end{tabular}
}
\caption{Different levels of reasoning using causal logic, with wireless examples.}\label{PearlCH}
\end{center}\vspace{-5mm}
\end{table*}
Next, we investigate how we can achieve the goals of explainability, generalizability, and energy efficiency using the fundamental building blocks of causality: \emph{Causal discovery, causal inference, and causal representation learning (CRL)}. 
\begin{definition}
\emph{Causal discovery} provides an algorithmic method to learn the inherent causal structure within observed wireless data, encompassing channel information, network topology, or any multi-modal source data. 
In contrast to statistical AI models, where the correlation between entities (features) within the data is inferred, a causal model accurately portrays the relationships among them that have an explanation (via directionality). Therefore, causal discovery offers an approach to design explainable AI-native wireless networks.
\end{definition}
One popular approach to represent a causal model through a graph is to use the language of \acp{SCM} \cite{PearlBasic2018} to describe the underlying mechanisms that drive a specific phenomenon of interest. \acp{SCM}, formally defined next, remain the most tractable way to perform causal logic, thus, hereinafter, we will be giving our examples via \acp{SCM}.

\begin{definition}
An \emph{SCM} is a collection of elements $<\mU, \mV, \mF, P(\bmU)>$, where $\mV$ represents endogenous variables (cause and effect variables), and $\mU$ represents exogenous variables (random, unknown noise). The structural functions $f_i \in \mF$ are designed so that $f_i(\textrm{PA}_i, \bmu_i)$ determines $\bmv_i$, where $\textrm{PA}_i \subset \mV$ and $\bmu_i \subset \mU$ specify the sets of parents (in the graph) and exogeneous inputs, respectively. The exogenous distribution $P(\bmu_i)$ determines the values of $\bmu_i$, and thus the distribution of endogenous variables $\mV.$ 
\end{definition}
Causal discovery assumed variables $\bmv_i$ to be random variables connected by a causal graph. However, real-world wireless observations are usually not structured into those variables. For example, high-dimensional channel information or raw source data that is either video or images or holograms. Herein, it is essential to exploit the concept of CRL. 
\begin{definition}
\emph{CRL} involves mapping large dimensional observations $\bmD\!=\!\left[d_1,\cdots,d_D\right]^T\!$ into a lower dimensional representation involving causal variables $\bmv_1,\cdots,\bmv_n$, captured as:
\beq
\bmv_i = f_i(\textrm{PA}_i,\bmu_i), \forall i.
\vspace{-1mm}\eeq
\end{definition}
We now turn our attention to leveraging the causal model for making informed decisions and drawing meaningful inferences through the concept of causal inference.
\begin{definition}
\emph{Causal inference} seeks to leverage the knowledge of the causal structure obtained from the causal discovery phase and the learned representations from CRL to make predictions and answer questions about the effects of interventions or counterfactuals over variables of interest. The utilization of causal inference thus empowers wireless systems with both inductive and deductive reasoning capabilities.
\end{definition}
Causal inference can be effectively implemented using two key concepts:
\begin{itemize}
\item 
\textbf{Interventions}: {Interventions can be defined as manipulating any causal variable in the graph using an exogenous random or deterministic process. The do-operator is a mathematical tool used to represent a physical intervention or manipulation of a variable in a causal model \cite{PearlBasic2018}.} In the context of a causal model represented as $Z \rightarrow X \rightarrow Y$, applying the do-operator to $X$ involves removing all incoming edges (that represent the generation of a particular variable from its descendants) to $X$ and setting to a specific value, $x_0$. This can be represented as ${do}(X) = x_0$.  For example, in Fig.~\ref{Causal_wireless_eg}, using a do-operator, the system designer can analyze the impact of different environments by considering ${do}(X)$ as implementing different propagation environments, e.g., Rayleigh, Rician, or purely \ac{LoS}.
{The set of interventions captured by $do(X)$}, encompassing a diverse range of functions that portray different forms of user or scatterer movements. These variations could span scenarios such as high-speed train travel or particle displacements within environments experiencing strong winds. Alternatively, {$do(X)$} could refer to random functions that produce novel combinations of scatterers. 
This simulation of intervention can be performed online, without any human intervention. Interventions allow us to study the causal effects of directly manipulating $X$ on the subsequent variables in the causal model, such as its effect on the outcome variable $Y$, while ignoring the causal relationships from $Z$ to $X$.
\item \textbf{Counterfactuals}: Counterfactual problems entail the process of reasoning about the causes behind events, envisioning the outcomes of different actions in hindsight, and determining which actions could have led to the desired results. 
In the previous example of the channel's impact on QoS, a potential counterfactual question could be: ``In a THz system, would the network's QoS have been more resilient to blockage if a \ac{RIS} were placed near the \ac{UE}/\ac{BS}, to alter the propagation environment?" Analytically, this counterfactual query can be formulated as optimizing the \ac{RIS} phase shifters:
\beq
\vspace{-0mm}\begin{aligned}
\left[\bphi^{\ast}\right] &= \arg\max\limits_{\bphi} \mbE\left[Y\mid \bmx_o,y_o\right], \\
\mbox{s.t.}\,\, & \norm{\mbE\left[Y\mid \bmx_o,y_o\right] - Y_{g}}^2 \leq \epsilon,
\end{aligned}\label{eq_counterfactual}
\vspace{-0mm}\eeq
where $\bmx_o$ is the set of observed channel parameters, $\bmy_o$ is the observed QoS (e.g., throughput or delay or reliability), and $\bphi$ is the vector of RIS phase shifts. $Y_g$ is the target \ac{QoS} value and $\epsilon$ is an arbitrary small quantity.
Hence, using counterfactuals, AI models can go beyond just observing the impact of an intervened environment. They can generalize their model behavior to hypothetical scenarios where fundamental changes to the network setup, such as introducing new technologies or altering configurations, could improve performance under various conditions. 
In traditional wireless problems, the objective is to find the optimal solution for a predefined objective function under specific constraints. In contrast, counterfactuals empower the AI system to adjust these objectives and constraints based on higher-level requirements, offering a more flexible and adaptive approach to problem-solving. Hence, the counterfactual formulation in \eqref{eq_counterfactual} outperforms conventional optimization approaches in wireless problems.
\end{itemize}

In Table.~\ref{PearlCH}, we provide several wireless examples that can capture interventional and counterfactual queries. In comparison to traditional deep learning or model-based optimization, a causal approach has major advantages. First, during training, causality allows training the \ac{AI} model on a diverse set of interventional and counterfactual queries, in addition to statistical observations. This approach thereby fosters \emph{distribution invariance}, to all interventional and counterfactual distributions. The second advantage involves the practical implications of understanding causal relationships within wireless networks for short-term and long-term planning. Certain network decisions, such as power allocation (for instance, transitioning from equal power allocation to an interference-aware water-filling algorithm), computational resource optimization (assigning downlink \ac{RAN} computations to either renewable or non-renewable energy nodes for sustainable wireless), or scheduling algorithm changes, can potentially be executed in real-time with minimal impact on \ac{QoE} performance. This is achievable because causal AI algorithms can compute minimal resource allocation or signal strategy changes by optimizing long-term average rewards (e.g., \ac{QoE}, given the network's current state). The resulting causal inference process also incorporates previous experiences and can be formulated as a counterfactual query, as demonstrated in \eqref{eq_counterfactual}. Conversely, long-term planning may involve addressing issues like blocked wireless links to specific \acp{UE}, which may require the introduction of \ac{RIS} on buildings as a potential solution. These longer-term initiatives may necessitate human intervention. 

Next, we delve into the limitations and challenges faced by current AI-native wireless networks. 

\vspace{-2mm}\section{Classical vs. Causal Reasoning-based AI-native Wireless Networks}
\label{Challenges}
This section outlines the prominent hurdles prevalent in contemporary AI-powered wireless systems, and the potential of causality to overcome them. 

 \vspace{-3mm}\subsection{Lack of Explainability and Trustworthiness}

 \subsubsection{Why are existing AI models not interpretable?} 
 In model-based Bayesian systems, the mapping from the data $X$ to $Y = f(X)$ is achieved by assuming a specific model, ${Y} = A(\btheta)X + N$, where $\btheta$ represents the set of estimated parameters. This is a common model for several wireless communication problems including channel estimation, data detection, and resource allocation \cite{TseCUP2005}. In Bayesian systems, explainability is evident since the model provides clear insights into which set of parameters contributes to a particular outcome. 
However, when dealing with non-linear underlying signal models that are difficult to accurately capture, deep learning systems often outperform Bayesian model-based systems. Despite their superior performance, 
these black-box models cannot comprehend the root causes behind deviations from particular performance objectives. 
As a result, these algorithms cannot discern the minimal configuration changes needed in terms of resource allocation or control signaling approaches to tackle performance issues effectively. This limitation impacts their capacity to autonomously restore the system to the designated target \ac{QoE}. This demands periodic retraining to sustain uninterrupted connectivity and ensure the expected \ac{QoE}. This can be exemplified using the THz blockage example discussed in Section.~\ref{Primer_why}.  

\subsubsection{Causal modeling for explainable wireless networks}
\label{Explainability}

\begin{figure}[t]
\centerline{\includegraphics[width=3.5in,height=2.6in]{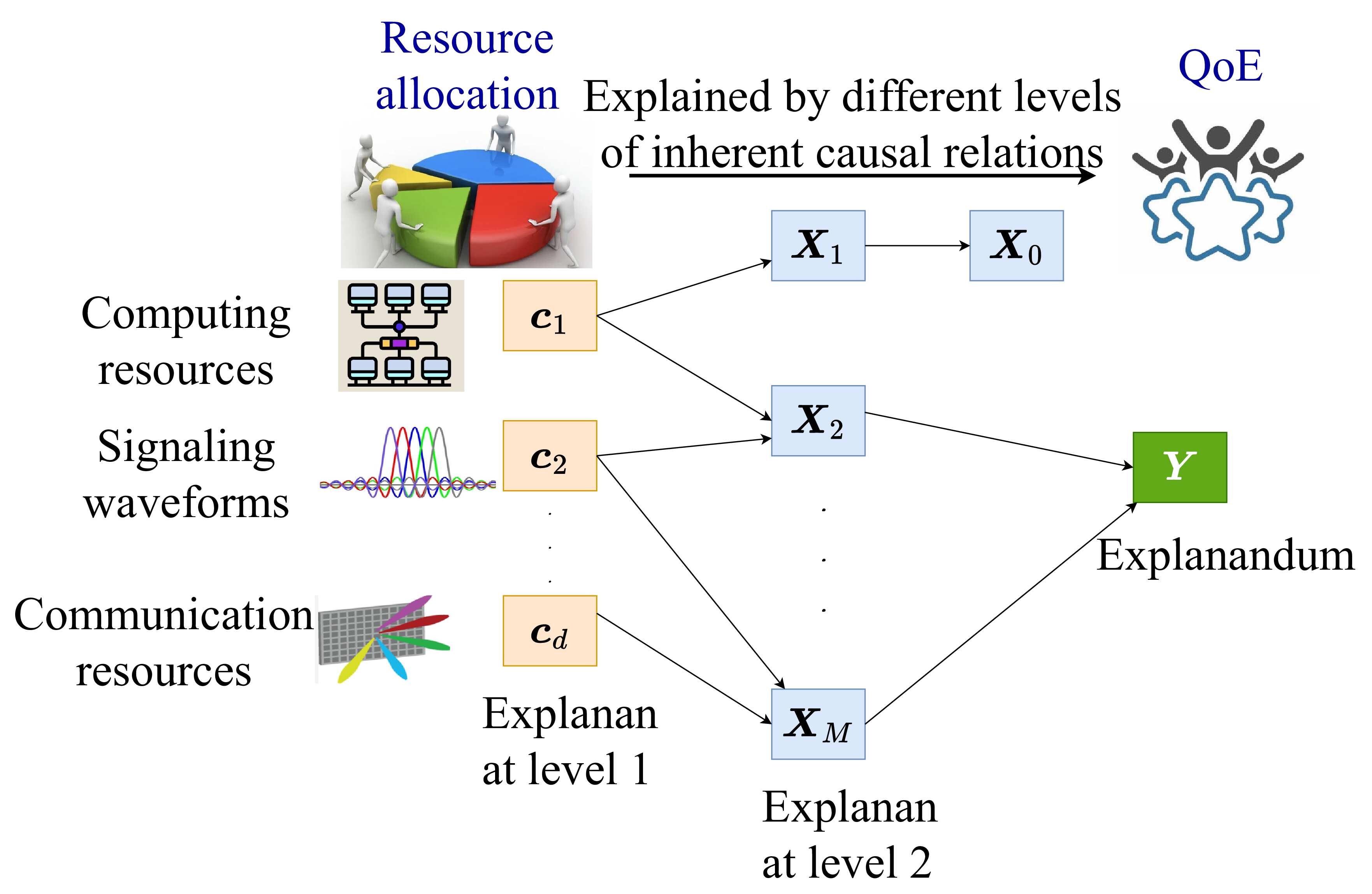}}\vspace{-0.5mm}
\caption{\small Structural causal model representing inherent causal relations among network variables.}
\label{SCM_Interventions}\vspace{-0mm}
\vspace{-7mm}
\end{figure}

To address the aforementioned challenge, we offer insights into the interpretability of causal models. 
For example, as shown in Fig.~\ref{SCM_Interventions}, in dynamic network slicing problems, variable $Y$ plays the role of a \ac{QoE} metric, which provides insights into user satisfaction and network performance based on resource allocation. Variables $\bmc_i$ represent control parameters like power or resource block allocation, signaling waveforms or computational resources for individual network slices. Each slice may have have its own varying \ac{KPI} requirements that must be met by the resource management framework. The intermediate variables could encompass network metrics such as \ac{SNR}, channel quality metrics, or average queue length at the transmitter. 
Variables $\bmX_0$ may represent miscellaneous control information irrelevant to the QoE. The objective is to compute the set of intervention variables $\mC_s \in \mC$ (defining all causal variables in the graph) and their corresponding intervention levels $\bmc_s$ to optimize the expected target outcome $\bmY$. Unlike traditional \ac{ML}-driven wireless solutions, leveraging a causal graph empowers the \ac{AI} algorithm to calculate optimal interventions, such as resource allocation variables, within a broader set $\mC$. This set may contain alterations to the generation of network variables compared to just utilizing observational data. Consequently, the outcome will be resource allocation algorithms that maintain invariance to these interventions as long as the causal relationships remain intact. 

A pivotal question emerges at this juncture: What renders the causal framework above interpretable? Herein, the explanation of any occurrence, referred to as the \emph{explanandum} \cite{MadumalAAAI2020} $\bmY$ (formally representing ``why did $\bmY$ occur?"), is accomplished through a sequence of causal relationships operating at various levels, along with the intervened values of the causal variables, called the \emph{explanans}. The level of detail in the explanation depends on the intended audience, which can vary from an end-user subscribing to the network service to a modem system engineer with in-depth knowledge of \ac{BS} infrastructure and signal processing. Explanations can then be tailored to either a micro-level, encompassing the entire sequence of detailed relations and processes, or a macro-level, which provides a high-level, natural language description of the process. Moreover, these explanations can be communicated offline or used autonomously online to provide various resource allocation and signaling scheme proposals to meet the expected business goals in intent-based wireless networks. An example of an online approach to leverage these explanations is to implement (causal) \ac{RL} with human feedback, akin to \cite{AiyappaArxiv2023}, but with explanations acting as the feedback component. {Here, causal Bayesian optimization (CBO) \cite{AgliettiPMLR2020} can be used to compute the optimal intervention set and their values.} CBO offers computational feasibility compared to traditional approaches like \ac{MCMC}. It operates akin to Bayesian inference but explicitly considers the optimization problem's causal structure. While this example primarily focuses on the causal relationships between resource allocation variables and \ac{QoE}, we stress that the CBO framework can be used for explainability in a broad range of wireless scenarios.

\vspace{-1mm}

\subsection{Inability to Reason and Generalize}
\label{Generalize}
Although explainability can be facilitated by the description of causal sequences underlying specific network behaviors, achieving a level of generalizability across multiple data distributions and domains (inductive reasoning) necessitates a distinct examination. We next discuss why traditional AI falls short here, and how causality can be leveraged for generalizability.

\subsubsection{Limited reasoning capacity of \acp{ANN}}
Generalization involves extending knowledge and capabilities across various domains, such as wireless environments (channels), network intents (energy consumption and end-to-end reliability), and control tasks (resource allocation and link adaptation). Current developments in AI-native 5G air interfaces as per 3GPP release 17 \cite{XLinArxiv2023}, primarily involves converting model-based signal processing approaches for beam management, channel estimation, and positioning into individual AI components. Most of these existing ML solutions for wireless networking \cite{ZhangCST2019} rely on ANNs, like transformers, \acp{CNN}, or \acp{RNN}, that lack distribution invariance. Hence, an \ac{ANN} approach consumes significant communication and computational resources due to the training overhead which would impede real-time immersive experiences, a crucial aspect of 6G. 

\subsubsection{Generalizability via causal reasoning}

Causal reasoning can allow generalizability, as defined next: Consider a fixed model class $M \in \mM$. Physically, $\mM$ may represent various wireless propagation environments or wireless control tasks. The model class $M$ encompasses various functions $f_i \in \mF$ (see the SCM definition in Section.~\ref{Primer}) that collectively form a specific observational distribution across various network variables $\mP_M$. Let $\mI$ be a set of interventions on the SCM, which is representative of external manipulations that bring about changes in environment or \ac{RAN} intent distributions. For the set of model classes $\mM$, we define distribution generalization as follows.
\begin{definition}
    Pair $(\mP_M, \mM)$ is considered to admit \emph{distribution generalization} to $\mI$ if, for any given $\epsilon > 0$, there exists a function $f^{\ast}_{\epsilon} \in \mF$ such that, for all models $ \widetilde{M}\in \mM$ where $\mP_{\widetilde{M}}=\mP_M$, the following condition holds:
 \end{definition}
 \beq
\abs{\sup_{i\in \mI} \left[\left(Y - f_{\epsilon}^{\ast}(X)\right)^2\right] - \inf_{f_{\diamond}\in \mF} \sup_{i\in \mI} \left[\left(Y - f_{\diamond}(X)\right)^2\right]}\leq \epsilon. \label{eq_distgen}
 \eeq

\eqref{eq_distgen} means that, for distribution generalization, one must only find an approximate solution $f_{\epsilon}^{\ast}(X)$ that is close to the minimax solution for all interventions on the wireless environment. Apart from distribution generalization, domain generalization can also be achieved through training across various interventional and counterfactual distributions, depending on how domains are defined. In some wireless contexts, domains can be characterized by differing contextual variables (with causal relations invariant), treatable as exogenous variables and can be manipulated through interventions. In contrast, the fundamental data domain varies in other cases (medical application vs. live sports event), making it challenging to address solely through interventions or counterfactuals. In contrast to conventional \ac{ML} approaches that require extensive training data from various random data distributions, causal ML leverages a single causal graph encompassing multiple data distributions generated using intervened \acp{SCM}, and accomplishes this efficiently with fewer samples.  Hence, this approach ensures the ultra-low latency, time criticality, and high accuracy synchronization needs of cyber-physical systems in 6G, while fostering seamless interactions between digital and physical realms. The training efficiency of causal \ac{ML} leads to energy-efficient algorithms, in contrast to other approaches to generalization such as transfer learning, meta-learning, and continual learning, as detailed next. 

\vspace{-2mm}\subsection{Energy Efficiency}

\subsubsection{Energy inefficient AI-native wireless networks} 

Traditional techniques for generalization such as foundation models for wireless networks yield \acp{NN} with a large number of parameters. For e.g., \acp{LLM} are being hailed as a game-changer for wireless networks. Particularly, they are being touted as solutions to the development of self-governing networks with intelligent decision-making at the edge \cite{BariahArxiv2023}.  
However, despite their promising potential, LLMs include billions of parameters which makes them energy-inefficient (during both training and inference) and unsustainable. 

\subsubsection{Causality as a tool for energy efficiency and sustainability}

Conventional methods aimed at enhancing energy efficiency, such as \ac{NN} pruning, can result in a reduction in performance. This is particularly undesirable for wireless networks given their stringent performance requirements highlighted in Section~\ref{Introduction}. Here, a more promising avenue is to leverage the capabilities of causal ML, to build lightweight foundation models. Next, we delve into the factors that could foster sustainability in wireless networking through causal \ac{ML}.
\begin{enumerate}
    \item \emph{Lighter models:} Causal models excel at simplifying complex relationships by focusing on essential causal factor, resulting in models that require less parameters, computations, and memory. By being lighter, these models consume less energy during both training and inference.
    \item \emph{Reduced training rounds and recurrent retraining:} Causal models efficiently capture stable causal relationships in wireless data across different contexts, requiring fewer training rounds and eliminating recurrent training. This reduction in training efforts significantly decreases the computational burden of large-scale models, enhancing overall computational efficiency.
    \item \emph{Reduced transmission:} By communicating just the causal states present in the data, one can reduce the amount of information transmitted, e.g., through causality-based \ac{SC}~\cite{ChaccourArxiv2022}, thereby decreasing communication energy consumption.
\end{enumerate}

Next, we turn our attention to practically realizing the above benefits using the distinct stages of causality outlined in Section~\ref{Primer}. 

\begin{figure}[t]
\centerline{\includegraphics[width=2.8in,height=2in]{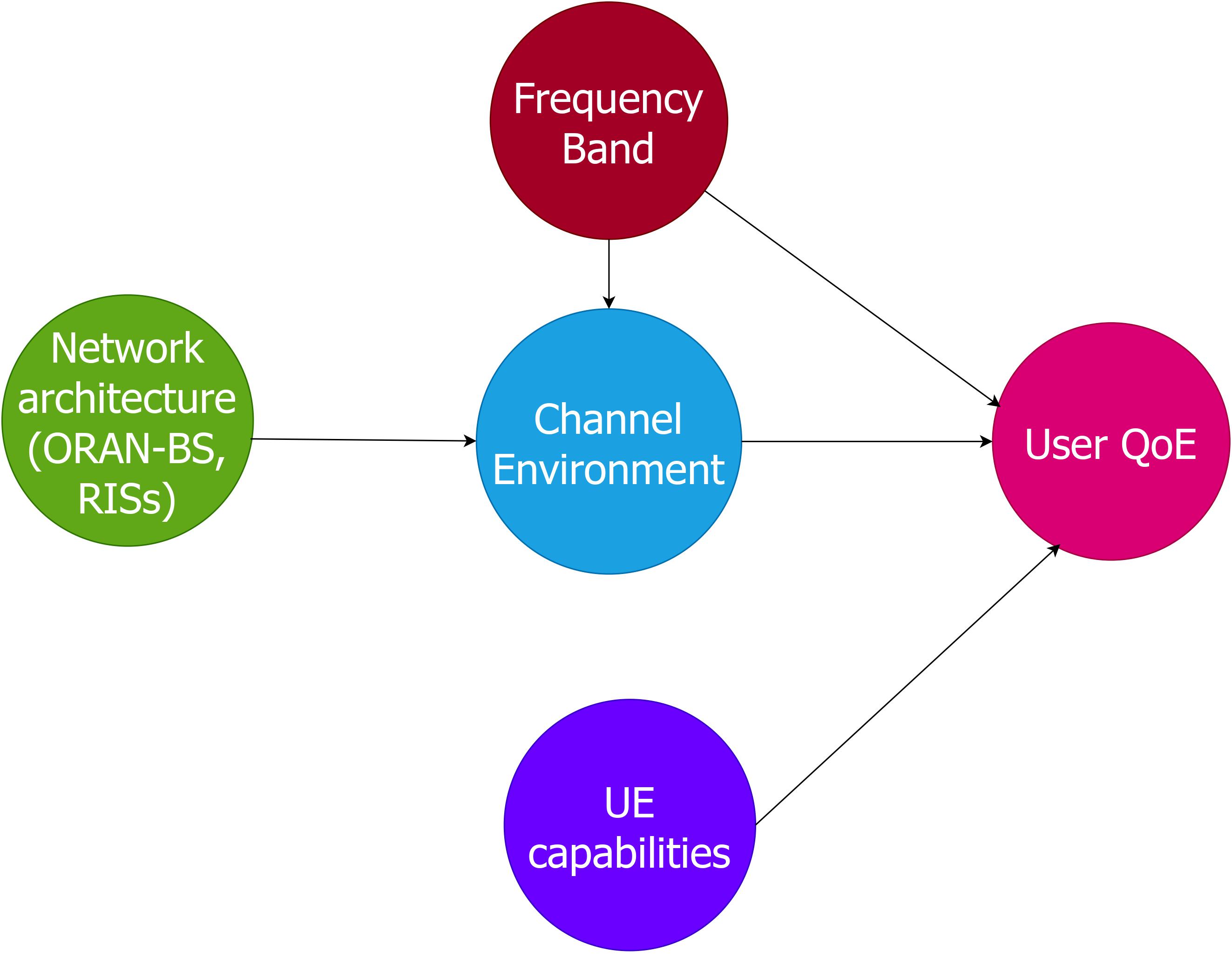}}\vspace{-1mm}
\caption{\small Simple example of a structural causal model.}
\label{CausalNetworks}\vspace{-0mm}
\vspace{-5mm}
\end{figure}\vspace{-1mm}

\vspace{-2mm}\section{Causal Discovery and Representation Learning for Next-Generation Wireless Networks}
\label{CausalDiscovery}

As explained in Section~\ref{Primer}, causal discovery involves learning the underlying causal graph representing the relations between different features present in the data (defined as \emph{causal variables}) that can directly or indirectly affect the behavior of a wireless system. However, it is crucial to identify the relevant causal variables before learning the causal graph. 
For example, consider the causal graph of Fig.~\ref{CausalNetworks}. Here, the user's \ac{QoE} is causally influenced by multiple factors such as the frequency band, channel, and the capabilities of the \ac{UE}, including \ac{ML} algorithms and processors. To isolate and analyze the specific impact of a causal variable, such as the frequency band, it is common to control or fix other causal dependencies (by do-calculus), like the channel environment and UE capabilities, at predetermined or set values. By doing so, one can observe and evaluate the direct causal effect of the frequency band on the user's QoE, without the confounding influence of other factors. 
Several pertinent problems in this domain warrant attention, and they are discussed next.

\vspace{-3mm}\subsection{Causal Graphical Models (CGMs) for Near-accurate Dynamic Wireless Environment Modeling}

We now discuss some open problems in wireless networks that can benefit from CGMs to achieve generalizable AI-native systems.

\subsubsection{Ultra-reliable beamforming 
for THz wireless} 
\label{BFTHz}

In THz wireless systems, ultra-massive \ac{MIMO} enables \ac{LoS} beamforming with multi-user interference cancellation under perfect \ac{CSI}. However, dynamic channel conditions due to user mobility or environmental changes require real-time adaptive algorithms. These algorithms fall into two categories \cite[Fig.~5]{ZhangCST2019}: Bayesian statistics-based methods (e.g., \ac{KF}) and data-intensive \ac{ML} methods (e.g., \ac{MAB} or \ac{RL}). An accurate user mobility model is essential to effectively use model-based algorithms like the \ac{KF}. 
Additionally, with an increasing number of antennas, the channel estimation process becomes more intricate, increasing the processing time. Consequently, a temporal gap emerges between the pilot transmission phase, dedicated to channel estimation, and the practical application of the estimated channel for downlink beamforming. During this interim period, the channel's characteristics may undergo substantial changes, exerting a notable influence on the precision of downlink beamforming. 
{To overcome the challenges of data-driven \ac{ML} models, a promising solution is to model the beam forming problem as a function of the causal dynamics (how the underlying \ac{CGM} evolves over time) of the channel components. This involves modeling the joint distribution of beamforming vector, and the channel, 
using variational causal networks \cite{ThomasICASSP2023}.} As demonstrated in \cite{ThomasICASSP2023}, the proposed causality-aware beamforming solution allows dynamic adaptability and outperforms existing beam-tracking solutions that struggle to model time-varying channels accurately. Moreover, the resulting network spectral efficiency brings the system closer to a perfect \ac{CSI}-based case. To model the causal dynamics, a promising approach is to use Granger causality \cite{LowePMLR2022}.


Granger causality provides two key advantages. First, it allows a generalization of the state space model that captures the channel variations across time to non-linear models. 
Second, training such causal reasoning enabled systems requires fewer samples for better generalization compared to conventional deep learning. This answers a key question: \emph{can causal \ac{ML} tools be leveraged for sample-efficient channel estimation in time-varying systems?} 
Sample efficiency can be achieved since enforcing causal constraints on the channel dynamics results in a reduced dimensional state space over which channel estimation happens. 



\subsubsection{Causal discovery for accurate physical twin modeling in digital twins}

The existing literature on physical twin modeling \cite{RuahArxiv2023} uses Bayesian inference. However, its performance is constrained by the specific data distribution used to construct the \ac{DT}, restricting applicability in diverse wireless environments. In contrast to this, causal discovery can be employed to model \acp{DT} of the entire physical network environment, including RAN architecture, edge servers, cloud services, and cell geography. The learned models of the environment dynamics, are sometimes
called as ``network state" models \cite{ChristoJSAITArxiv2022}. 
A network \ac{DT} located at a cloud would replicate
various latent network features, including signals, coverage, interference, traffic behavior, and
user mobility, across different \ac{OSI} layers. This leads to better prediction of network actions, including resource allocation decisions or autonomous tasks relevant to the physical system modeled by the \ac{DT}. However, learning such models purely from observations will be biased due to the presence of confounding variables, which may be unknown to the \ac{BS} or the cloud hosting the \ac{DT}. 

Contrary to the literature, \acp{CGM} offers a better understanding of causal relationships among network variables, enhancing model precision (explainable models, as discussed in Section.~\ref{Explainability}) and network traffic prediction. This improves network optimization, resource allocation, and decision-making, boosting wireless network performance and efficiency. Real-time reconstruction and prediction via \ac{DT} models are crucial for \ac{CI} applications, offering sample efficiency and reduced training efforts for real-time network control and tasks across diverse applications.

\vspace{-0mm}\subsubsection{\acp{CGM} for wireless training data generation}
\label{Section_CGM}


A scarcity of training data can adversely impact the performance of AI-native network models that make real-time control decisions. 
Existing schemes include using score-based models and \acp{GAN} to learn the posterior distribution of either the wireless channels or user locations \cite{StuderAccess2018}, and 
 using noisy uplink pilot signals as the observations.  However, these generative modeling schemes cannot fully exploit the underlying structure present in the data, particularly for high-frequency wireless systems, where the channel data exhibits high levels of structure. 
 To overcome these challenges, utilizing the learned causal graph $\mG$, makes it easier to create novel combinations of training data by interventions and counterfactuals as discussed in Section.~\ref{Primer}. 
The resulting generative AI algorithms are called \emph{causal generative models}. 
Exploiting training data augmentation include 1) generating diverse network scenarios, including those arising from natural disasters, congested network traffic, software malfunctions, and hardware imperfections, 2) creating data that accurately reflects the complexity of large-scale wireless network deployments, considering interactions among numerous devices and users, particularly when employing \ac{DT} models, and 3) developing methods for generating a diverse set of tasks suitable for applications of \ac{CI} over wireless networks.


\vspace{-2mm}\subsection{Resilient Wireless Networks}

Resilience encapsulates a wireless system's ability to sustain functionality in the presence of errors, adapt to erroneous influences, and promptly recuperate its normal functionality. Broadly, the comprehensive notion of resilience encompasses various system attributes, namely \emph{detection, remediation, and recovery} \cite{ReifertTVT2023}.  
Only few studies, e.g., \cite{ReifertTVT2023}, and \cite{NaderiAlizadehTSP2023} attempted to define resilience for wireless networks. 
These state-of-the-art methods rely on model-based non-convex optimization algorithms to build resilient communication systems. However, model-based algorithms are not effective when the dynamics of the wireless or \ac{RAN} environment diverge from the assumptions inherent in the model, as explained in Section~\ref{Primer}. In contrast, we propose a framework founded on causality principles aimed at revolutionizing the design of resilient wireless systems. Consider that as outlined in Fig.\ref{SCM_Interventions}, \ac{BS} powered by a \ac{DT} is aware of the causal graph $\mG$ that represents the intricate causal relations among network variables. Using $\mG$, the \ac{BS} can continuously monitor future \ac{QoE} values while the wireless network is in a particular state $\bms_t$. While monitoring, the \ac{BS} can detect prospective \ac{QoE} violations in the near future. This can be formulated as computing the average \ac{QoE} over a time duration $T$, given that the \ac{BS} follows given resource allocation policy $\pi_t(\bma_t\mid\bms_t)$:
\beq
\begin{aligned}
\bar{Y} & = \mbE_{p(Y\mid \bms_t, do(\bma_t\rightarrow \bma_{t+T}))} \left[Y\right], \\
\mbox{s.t.}\,\,\, &\bma_t \in \mA.
\end{aligned}
\eeq
If $\bar{Y}$ is below a specific threshold, the network proceeds to the remediation phase, i.e., it solves the following counterfactual optimization:
\beq\begin{aligned}
\argmax\limits_{\pi(\bma_t\mid\bms_t)} \,&\mbE_{Y\sim \mD} \left[Y\right], \\
\mbox{s.t.}\,\,\, &\bma_t \in \mA,
\end{aligned}
\eeq
where $\mA$ represents the space of feasible resource allocation variables capturing the system constraints. Here, $\mD = \mD_o \bigcup \mD_I$, where $\mD_o$ represents the observational dataset corresponding to the state transitions from time $t$ to $t+T$, given the current policy $\pi_t(\bma_t\mid\bms_t)$, and $\mD_I$ represents the interventional dataset. Here, various interventional distributions can be generated using causal generative modeling. Hence, the above detailed causal inference steps help to achieve the resilience goal in next-generation wireless networks.
Thus far, we examined how various signal processing functions (joint computing, and control aspects) throughout the \ac{OSI} layers can benefit from causal discovery. Next, we discuss how, causal discovery can contribute to transmission reduction (communication aspect).

\vspace{-4mm}\subsection{Causal Discovery and Representation Learning for Semantic Communications}

{In SC systems, a rigorous way to define the semantics at the transmit side is as the inherent causal structure within the data \cite{ChaccourArxiv2022}, and \cite{ChristoTWCArxiv2022}. 
The causal graph  among source data features can be learned as the posterior distribution of the entities under different interventions in the task distribution (in short, a multi-modal distribution compared to variational techniques \cite{HoffmanJMLR2013}) using techniques like generative flow networks \cite{DeleuArxiv2022}. 
After learning the causal graph, CRL can be used to form a unique representation for states that share comparable cause-and-effect repertoires. This leads to a reduction in the amount of information conveyed compared to systems that do not employ CRL. 

In classical information theory, mutual information is symmetric, meaning that, for any two random variables $X$ and $Y$, $I(X;Y) = I(Y;X)$, and it measures the information conveyed by the association between two random variables. However, in the context of causality-based SC systems, the association between two variables is directed, emphasizing the causal relationships rather than mere information exchange. 
A promising alternative here to capture the semantic information is the \ac{IIT} \cite{ChristoJSAITArxiv2022}. 
 Building on the aforementioned concepts of semantic representation and semantic information rooted in causality, we identify at least two key open problems:
 \begin{itemize}
 \item Formulating robust transmission schemes for multi-user systems with diverse semantic languages and varying \ac{QoE} requirements.
\item Uncovering causal relationships in control information across network layers and designing scheduling algorithms using semantic representations obtained from the source data.
\end{itemize}
These challenges can be effectively addressed through a fusion of causality and hypergame theory \cite{KovachGT2015} or by incorporating theory of mind reasoning \cite{YuanArxiv2021}.}

\subsection{Causal Discovery for Integrated Sensing and Communication (ISAC)}
\label{ISAC_1}

The ISAC paradigm synergistically combines sensing and communication functions, potentially sharing common spectral, signaling, and hardware resources~\cite{chaccour2023joint}. ISAC offers two advantages: \emph{integration gain}, achieved by enhancing resource efficiency, and \emph{coordination gain}, which involves co-design efforts to balance network performance (user \ac{QoE}) or attain sensing benefits (higher sensing resolution). Some applications of ISAC include 1) multipath parameter estimation from nonlinear channel models (see Section.~\ref{Primer}) for simultaneous localization and mapping (SLAM) of the environment, 2) integration of 
 sensing (using, e.g., LiDAR, radar, cameras) with communication to enable vehicular communication for safer and more efficient transportation, and 3) advanced immersive experiences within a metaverse by sensing the user's physical location, movements, and gestures and communicating this information to enhance interactions with the metaverse environment. ISAC applications require real-time monitoring, synchronization, and the processing of large amounts of distributed sensing data across multiple devices to leverage the benefits mentioned above. This leads to challenges that existing AI tools cannot handle, as explained next.

\makeatletter
\renewcommand{\@IEEEsectpunct}{\ \,}
\makeatother\subsubsection{Why is data-driven AI insufficient for achieving the integration and cooordination gains of ISAC?}

\vspace{-0mm}\begin{itemize}

\item \emph{Dynamic target tracking:} Accurately tracking objects in the wireless environment, especially for SLAM, is challenging due to the mobility of scatterers or users. Dynamic beam alignment and tracking face additional challenges, especially with high mobility. 
{Since data-driven AI struggles to capture the time-varying dynamics of nonlinear signal models accurately, this leads to beam misalignment and jeopardizes user QoE.}
\item \emph{Resource efficiency in distributed sensing:} 
Utilizing joint transmissions from multiple sensing nodes improves sensing capabilities, and processing communication signals from distributed receivers enhances sensing performance. Designing such distributed systems is challenging, involving considerations like synchronization, control overhead, and complex issues related to processing functions.
Moreover, resource efficiency, through the integration gain becomes challenging due to the substantial volume of sensing information that must be communicated (e.g., metaverse scenarios). Advanced distributed learning models like transfer learning, federated learning, and graph neural networks could facilitate the efficient integration of information collected from various locations. These techniques also aid in compressing distributed sensing data and reducing coordination overhead. {However, the data-driven AI solutions impede real-time synchronization and lack scalability as the network size fluctuates dynamically.} This is attributed to changes in the communication context as the number of sensing nodes varies, which can be viewed as shifts in the data domain.

\end{itemize}

\makeatletter
\renewcommand{\@IEEEsectpunct}{\ \,}
\makeatother
\subsubsection{How can causal discovery enhance transmission efficiency, and optimize real-time ISAC systems?}

To efficiently address challenges in dynamic target tracking and beam alignment, the Granger causality concept, as detailed in Section~\ref{BFTHz}, can be applied to construct an almost accurate model of the time-varying target object or \ac{UE} device. For beam tracking, causal generative modeling can be employed to make precise predictions of users' or scatterers' positions. Beam tracking can be improved further by using nonlinear minimum mean squared error (MMSE) methods, considering predictions as prior information. The choice of nonlinear MMSE is justified due to the nonlinear nature of signal models (beamforming vector as a function of user and scatterer positions, observed by sensors). The nonlinear MMSE method involves computing posterior distributions of beams by incorporating the causal dynamics described in Section~\ref{BFTHz}.

Efficient communication-aided distributed sensing can be achieved through the utilization of causal reasoning games, \cite{HammondAI2023}. Here, each sensing node can learn the causal graph, represented as an \ac{SCM}, that captures the causal relations among different network decision variables (power, spectrum, beamforming vectors, ISAC waveforms). By employing counterfactual reasoning, each sensing node can estimate other users' communication strategies and data distributions, Additionally, each sensing node can optimize communication and sensing strategies by leveraging game-theoretic techniques and the estimated beliefs about other nodes. Hence, causal discovery using SCMs and game-theoretic tools allows sensing nodes to make informed decisions about what to communicate. This dynamic process enables also the optimization of spectrum and time resources (how to communicate). Consequently, causal discovery plays a pivotal role in improving resource efficiency within distributed sensing scenarios, enabling the fulfillment of the requirements for dynamic adaptability and real-time synchronization essential for next-generation wireless networks.

Next, we answer a crucial question: How can we effectively use the learned causal graph to make informed control decisions and draw meaningful inferences for multiple tasks in real-time monitoring of autonomous wireless networks?  The key to addressing this lies in causal inference.

\section{Causal Inference for Next-Generation Wireless Networks}
\label{CausalInference}
Causal inference facilitates the execution of interventional and counterfactual reasoning across various wireless tasks. We identify specific wireless challenges where causal inference can contribute to achieving future wireless networking goals such as time criticality, intent management, and dynamic adaptability, as illustrated in Fig.~\ref{CausalAI_Vision}.

\vspace{-2mm}\subsection{Causal Inference for Wireless Control}

The radio resource management problem is typically non-convex and becomes computationally complex, particularly for large networks. Moreover, in dynamic wireless networks, model-based signal processing algorithms may be unable to maintain consistent performance across various scenarios. In this context, \ac{RL} has been a popular approach for wireless control problems. However, it is a passive approach that lacks reasoning capabilities and relies solely on observational data, typically presented as a set of state-action-reward tuples. 
We next describe the key limitations of \ac{RL} algorithms such as Q-learning, SARSA (state-action-reward-state-action), and deep Q-Networks (DQNs).
\makeatletter
\renewcommand{\@IEEEsectpunct}{\ \,}
\makeatother
\subsubsection{Why is RL using observational data alone insufficient for addressing wireless control problems?}
\label{RL_challenges}


Wireless resource allocation problems are often posed as partially-observable Markov decision processes (POMDPs) \cite{FerianiComST2021}. 
Traditional RL methods for addressing such problems are limited by their reliance on specific observation and state transition distributions. 
Furthermore, deep Q-learning-based \acp{NN} lack interpretability, making it challenging to explain the rationale behind a specific action in response to a given state. 
Thus, in this realm of \ac{RL} for wireless networks, a fundamental question emerges: \emph{can offline data obtained from UL or DL observations be combined with online data from experimentation (via \ac{DT} model simulations or causal generative models) to enhance the performance of a learning agent?} Previous studies \cite{KasgariTCOM2020} addressed this problem using GANs, however, \acp{GAN} primarily learns the posterior distribution of the data based on the provided dataset and hence they are not generalizable. We propose to overcome these major limitations by designing a causal RL framework which is an instance of causal inference, as discussed next.
\makeatletter
\renewcommand{\@IEEEsectpunct}{:\ \,}
\makeatother
\subsubsection{Causal RL for real-time resource management}

\ac{RL} using causal inference involves two key elements: \emph{experimentation and observation}. In decision problems like POMDPs, causal queries play a natural role where actions directly correspond to interventions. One such example is determining the causal effect of an action (intervention) on future rewards based on past information about the POMDP process. We can evaluate the causal query as $p_s(\bms_{t+1}\mid \bms_{0\rightarrow t},{do}(\bma_{0\rightarrow t}))$, where ${do}(\bma_{0\rightarrow t})$ can be generated by different possible distributions of $\bma_t$. By formulating decision problems as causal queries, we can explore how actions influence outcomes and make informed decisions based on their expected causal effects. Here, a particular set of causal queries may represent a specific task or a wireless environment.
 For creating an \ac{NN} model that can generalize across various wireless environments or tasks, the accumulated learning data obtained through causal queries can be distilled into an \ac{NN} using \emph{``behavioral cloning"}\cite{BainMI1995}. The learning history, which encompasses the causal state transitions, can be defined for a specific task or wireless environment $n$ as:
\beq
h_t^{(n)} = \left(\bmo_0,\bma_0,\bmr_0,\bmo_{1},\cdots,\bmo_T,\bma_T,\bmr_T,\bmo_{T+1}\right)_n,
\eeq 
where $\bmo_t$ represents the observation.
Assume that the lifetimes of the agents, called learning histories, are generated by the source \ac{RL} algorithm for a set of individual tasks or wireless environments $\{\mM_n\}_{n=1}^N$. This process creates a dataset $\mD = \{(h_t^{(n})\sim P_{\mM_n}\}$.
Next, by \ac{AD} \cite{LaskinArxiv2022}, we distill the behavior of the source algorithm into a sequence model that maps long histories to probabilities over actions. \ac{AD} includes training \ac{NN} models $P_{\btheta}$ with parameters $\btheta$ using a negative log likelihood (NLL) loss. The \ac{NN} parameters can be obtained by minimizing the following loss function:
\beq
\mL(\btheta) = \sum_{n=1}^N\sum_{t=1}^T \log P_{\btheta}(\bmA = \bma_t^{(n)}\mid \bmh_t^{(n-1)},\bmo_t^{(n)}).
\eeq

In scenarios involving multiple wireless tasks, this approach guarantees that the learned policy remains task-agnostic. When the source algorithm is trained across various wireless environments and subsequently distilled into a \ac{NN}, it results in the development of a \ac{NN} that is invariant to environment changes. In summary, training across a wide range of environments or tasks and then distilling this knowledge into a sequence model leads to the development of \emph{causal foundation models}. Nevertheless, in hierarchical and distributed wireless networks, due to the intrinsic lack or uncertainty of available information (that may include channel state, computational, and communication resources available to interfering users) at each network node, we need to look beyond causal \ac{RL}. Herein, a possible solution is the framework of \acp{MAB}.



\subsubsection{Causal bandit problem for real-time decision making in wireless networks}

  \ac{MAB} represents a category of sequential optimization problems wherein, given a collection of arms (actions), a player selects an arm in each round to obtain a reward. Traditional \ac{MAB} algorithms cannot satisfy the ultra-synchronization demands of 6G wireless systems because of their inability to maintain distribution invariance and their substantial need for extensive training data. In this context, possessing causal knowledge in the form of an \ac{SCM} $\mathcal{M} = (\mV, \mE, \mF, P(\mE))$ enables wireless agents to acquire knowledge about the distribution of non-intervened variables, $p(\bmv^n\mid do(\zeta)=I)$, where $\zeta$ represents the action (arm) taken, $I$ captures the intervened values, and $\bmv^n \in \mV$. 
The objective is to minimize the cumulative regret with respect to $Y \in \mV$, and captured as $\mathcal{R} = \sum\limits_{n=1}^{T}(Y^n - \max_{\mathbf{\zeta}}\mathbb{E}[Y|\mathbf{I} = \mathbf{\zeta}])$. The expectation operator $\mbE$ is applied to the policy distribution (over arms) and the non-intervened variables, which distinguishes it from traditional \ac{MAB} approaches where only the pairs $(\zeta,Y)$ are known. Consequently, additional knowledge in the form of an \ac{SCM} ensures distribution invariance without requiring additional training data as is the case with conventional \ac{MAB}.

\acp{MAB} pose more significant challenges in multi-user scenarios, such as opportunistic spectrum access within a metaverse, where users need to collaborate to assess the \ac{KPI} needs of other users. Herein, it is essential to consider not only the causal relationships within the environment but also the inter dependencies among users' strategies. 
A promising approach to solve such multi-user \ac{MAB} problems is to consider a merger of causality and game theory \cite{HammondAI2023}.

\subsection{Causal Inference for Intent-based Wireless Networks}

Intent-based wireless networks aim to translate human intentions (called business intent) into network configurations through automation and AI, improving network management, efficiency, and responsiveness. One example of business intent involves optimizing wireless network sustainability while ensuring a consistent \ac{QoE} for end-users. 
Despite industry interest \cite{GomesICT2021}, the wireless literature lacks a comprehensive articulation of intent-based networks. Here, we explore how causal inference is a promising approach for building intent-based networks, characterized by the following operations forming a closed loop.
\vspace{-1mm}\begin{itemize}
\item \emph{Intent assurance:} Once the service \acp{KPI} are defined, \emph{measurement agents} consistently oversee the network variables, while \emph{assurance agents} assess the network's compliance with its intended objectives. 
\item \emph{Corrective actions:} If the business goals have not been met, the node-level \ac{ML} agents, known as \emph{proposal and policy evaluator agents}, are consulted to offer predictions and recommendations. A conventional approach would involve using deep RL agents for both proposal (RL actions) and evaluation (reward computation) tasks. However, as discussed in Section.~\ref{CausalDiscovery}, the data-driven \ac{RL} challenges hinder the autonomous intent management functionality.
\end{itemize}

These closed-loop operations pave the way for constructing a fully autonomous AI-native wireless network, provided that the underlying AI agents do not require recurrent training and the network decisions are interpretable, a requirement that classical AI solutions often cannot fulfill. In contrast, performing causal inference using a causal graph that describes the relations among network decision variables, which may span various \ac{OSI} layers, offers the following benefits: 
\begin{itemize}
\item \emph{Performing self-reflection or self-critique:} As discussed in \cite{ParkArxiv2023}, this approach is only feasible if the \ac{ML} model is interpretable and explainable, which causality provides.
\item \emph{Decomposing intents into subgoals and performing deductive reasoning:} This is achieved by decomposing a larger \ac{SCM} into smaller SCMs. Splitting SCMs is advantageous as decisions at higher network layers may have ramifications across lower layers. It facilitates distributed interventions or counterfactuals. Combining causal effects from individual SCMs allows deductive reasoning, offering novel insights into network decisions across different \ac{OSI} layers.
\end{itemize}

One promising approach to realize the above benefits is to compute parallel policies for each subgoal using either multi-objective \ac{RL} (MO-RL) \cite{AbdolmalekiArxiv2021} or causal reasoning games \cite{HammondAI2023}. After obtaining subgoal policies, the evaluator agent can use supervised learning to distill these
distributions into a new parameterized policy.
 These recommendations help execute autonomous actions through a closed-loop system that incorporates advanced analytics, automation, and \ac{AI}. This autonomous execution ensures that the network operates efficiently, adapts to changing conditions, and achieves the desired service KPIs without the need for manual intervention.

\subsection{Causal Inference for Integrated Sensing and Communication (ISAC)}
In addition to the challenges related to resource efficiency and dynamic target tracking, as discussed in Section~\ref{ISAC_1}, ISAC applications also require effective management of non-linear signal models and joint waveform design for sensing and communication functionalities. Furthermore, they must enable dynamic resource sharing between sensing and communication functions to leverage the benefits of integration and coordination gains, as elaborated next.
\subsubsection{Why is data-driven AI inadequate for achieving the integration and coordination gains of ISAC?}

\vspace{-0mm}\begin{itemize}

\item \emph{Nonlinear signal processing:} ISAC functions may require full-duplex operation where, 
the transmitter might be actively transmitting the ISAC waveform while, concurrently, the receiver captures the back-scattered or reflected signals for subsequent radar processing. Efficiently canceling self-interference (SI) at the antenna, analog, or digital domains requires a substantial understanding of circuit non-linear models, antenna coupling, and time-varying channel characteristics. However, approximating such time-varying SI using data-driven AI need not result in models that are accurate across time, and the recurrent retraining required introduces significant overhead.

\item \emph{ISAC waveform design:} ISAC systems allows using a single waveform for both radar and communication purposes. In such cases, the joint waveform must be rigorously designed to fulfill communication goals like data rates, latency, and \ac{QoE}, all while preserving the desired sensing capabilities. This becomes particularly challenging when: 1) dealing with different requirements for communication and sensing systems, like peak-to-average power (PAPR) constraints for \ac{OFDM} systems and signal quality for sensing, 2) accounting for hardware imperfections and non-linearities, and 3) managing highly-mobile scenarios where radar and communication channels change rapidly. Classical AI struggles to adapt to dynamic (non-stationary) channels and non-linear signal models that are specific to different devices or domains. 
Because of the conflicting objectives and the large number of optimization variables related to communication and sensing, classical AI solutions may not achieve near-accurate Pareto-optimal solutions \cite{AbdolmalekiArxiv2021} without extensive training. 

\item \emph{Resource sharing between sensing and communication in dynamic environments:}  Allocating wireless resources like time, frequency, and power efficiently between sensing and communication is challenging due to tradeoffs between sensing quality and communication goals. ML can enhance resource allocation by using prior observations and context to optimize decisions. For instance, \acp{RNN} can proactively predict user requests and \ac{RL} can be used to explore resource optimization options. Nevertheless, \ac{RNN} cannot handle large input sequences, such as extended user request histories or contexts. Additionally, \ac{RL} is often constrained by the dataset it was trained on, limiting its adaptability to evolving, non-stationary conditions or novel scenarios. 
\end{itemize}

\makeatletter
\renewcommand{\@IEEEsectpunct}{\ \,}
\makeatother
\subsubsection{How can causal inference address nonlinearity, joint waveform design, and optimize real-time resource sharing in ISAC systems?}

To address the inherent nonlinearity of SI signals, a promising strategy involves leveraging causal discovery to gain insights into the underlying ``physics" governing the data. The generation of these signals, influenced by hardware imperfections and the SI channel, is not random; it is guided by specific physical principles (electromagnetic as well as circuit theory-based). By constructing a causal graph that captures relationships among channel parameters, electronics hardware characteristics, and antenna coupling coefficients, the causal model can be extended to accommodate various interventional and counterfactual distributions of channels, frequency bands, or antenna characteristics.

Causal MO-RL can efficiently handle multi-objective optimization in ISAC waveform design. For example, consider that sensing and communication signals occupy different subcarriers within an OFDM symbol under a multi-antenna transmitter. In this scenario, the optimization variables could include finding the optimal number of subcarriers, the complex coefficients for the sensing subcarriers, and the beampatterns for radar and communication signals. Radar signals may have a wider beamwidth requirement (due to omnipresent scatterers) in contrast to communication users' narrower beam requirements. These variables are computed with the dual objective of minimizing the target detection error (e.g., delay and Doppler mean-squared error) at the sensing receiver and maximizing the user's QoE. The multi-objective optimization has to be performed under several constraints that may include transmit power and PAPR. By integrating causal inference into MO-RL problems, it becomes possible to identify the minimal interventions needed to optimize specific objectives. The interventions here involve the number of subcarriers or the sensing signal characteristics, which depend on exogenous variables such as environment characteristics (e.g., scatterer parameters as discussed in Section.~\ref{Primer}) that are time-varying and the allocated spatial or spectrum resources. Causal inference simplifies the solution space by estimating the minimal required interventions to be performed. Hence, it allows computation of near Pareto-optimal solutions, surpassing the capabilities of data-driven MO-RL that may be stuck in a local optima. Here, causal inference can significantly improve the overall performance of ISAC systems compared to generic waveforms that do not consider the causal dependency of user \ac{QoE} or sensing quality on factors like environment geometry, user location, mobility patterns, and hardware imperfections. To create resource-sharing algorithms that are both generalizable and explainable, one can explore causal inference techniques such as causal \ac{RL} and causal \ac{MAB}.

\vspace{-2mm}\section{Conclusion and Recommendations}

This article laid out a bold new vision for designing AI-native wireless systems (6G and beyond) founded on the principles of causal reasoning. 
We conclude with \emph{three key recommendations}:
\begin{itemize}
\item \textbf{Causal Reasoning as a Bedrock for AI-Native Wireless Networks:} Designing AI-native wireless networks is not going to be successful if it relies simply on reusing data-driven AI tools to mimic existing wireless functions. Instead, there is a need to develop novel AI frameworks that incorporate the desirable properties of explainability, reasoning, generalizability, and sustainability. In this regard, we recommend embracing causal reasoning as the cornerstone of AI-native wireless systems as it enables these key characteristics. {Causal reasoning can be integrated across the networking stack, including 1) source data sampling (selecting causal variables for extraction), 2) \ac{QoS} management, and cell selection procedures in network layer using causal structure among multi-user channels, 3) data encoding (using causal representations), 4) scheduling (incorporating causality awareness into user data and KPIs for resource management), and 5) communication (users seeking relevant interventions or counterfactuals to gain insights about network attributes).}

\item \textbf{Refined Models with Causality to Mitigate Bias:} We advocate transitioning away from excessive reliance on training data, characteristic of data-driven networking with foundation models in general. Instead, we emphasize the importance of refining models and features using causality to mitigate potential biases within AI frameworks. In other words, we recommend designing causality-based foundation models that can lead to interpretable and lighter \ac{ML} models, which are crucial for achieving next-generation wireless network goals as shown in Fig.~\ref{CausalAI_Vision}. 
\item \textbf{Scalable AI-Native Wireless Networks using Layer-wise SCMs:} Causal \ac{ML} comes with its own set of challenges. The profusion of network variables spanning multiple wireless networking layers can result in intricate and expanding causal relationships, particularly as the network grows in scale. In this context, we recommend devising a path to scale causality-aware AI-native wireless networks without disrupting existing 3GPP architectures. A practical strategy to address this scalability challenge involves constructing layer-specific \acp{SCM} and devising causality-aware algorithms tailored to specific \ac{OSI} layers. 
This distributed architecture can also align with the open RAN architecture of 5G and beyond, enabling causal reasoning across individual layers distributed across different network components or implemented as virtualized network functions.
\end{itemize}

\bibliographystyle{IEEEbib}
\def\baselinestretch{0.9}
\bibliography{bibliography,semantics_refs}

\end{document}